\documentclass[twocolumn,showpacs,preprintnumbers]{revtex4}
\usepackage{graphicx}
\usepackage{dcolumn}
\usepackage{bm}
\begin{document}
\newcommand{\ds}{\displaystyle}
\newcommand{\be}{\begin{equation}}
\newcommand{\en}{\end{equation}}
\newcommand{\bea}{\begin{eqnarray}}
\newcommand{\ena}{\end{eqnarray}}
\title{Evolving Lorentzian wormholes supported by phantom matter and cosmological constant}
\author{Mauricio Cataldo}
\altaffiliation{mcataldo@ubiobio.cl} \affiliation{Departamento de
F\'\i sica, Facultad de Ciencias, Universidad del B\'\i o--B\'\i
o, Avenida Collao 1202, Casilla 5-C, Concepci\'on, Chile.}
\author{Sergio del Campo}
\altaffiliation{sdelcamp@ucv.cl} \affiliation{Instituto de F\'\i
sica, Facultad de Ciencias, Pontificia Universidad Cat\'olica de
Valpara\'\i so, \\ Avenida Brasil 2950, Valpara\'\i so, Chile.}
\author{Paul Minning}
\altaffiliation{pminning@udec.cl} \affiliation{Departamento de
F\'\i sica, Facultad de Ciencias F\'\i sicas y Matem\'aticas,
Universidad de Concepci\'on, Casilla 160-C, Concepci\'on, Chile.}
\author{Patricio Salgado}
\altaffiliation{pasalgad@udec.cl} \affiliation{Departamento de
F\'\i sica, Facultad de Ciencias F\'\i sicas y Matem\'aticas,
Universidad de Concepci\'on, Casilla 160-C, Concepci\'on, Chile.}
\date{\today}
\begin{abstract}
In this paper we study the possibility of sustaining an evolving
wormhole via exotic matter made of phantom energy in the presence
of a cosmological constant. We derive analytical evolving wormhole
geometries by supposing that the radial tension of the phantom
matter, which is negative to the radial pressure, and the pressure
measured in the tangential directions have barotropic equations of
state with constant state parameters. In this case the presence of
a cosmological constant ensures accelerated expansion of the
wormhole configurations. More specifically, for positive
cosmological constant we have wormholes which expand forever and,
for negative cosmological constant we have wormholes which expand
to a maximum value and then recolapse. At spatial infinity the
energy density and the pressures of the anisotropic phantom matter
threading the wormholes vanish; thus these evolving wormholes are
asymptotically vacuum $\Lambda$-Friedmann models with either open
or closed or flat topologies.

\vspace{0.5cm} 
\end{abstract}
\smallskip\
\maketitle \preprint{APS/123-QED}
\section{Introduction}
It is well known that in 1917 Einstein added a constant term into
his field equations of general relativity allowing static
cosmological models for the Universe. This term, called
cosmological constant, would create a repulsive gravitational
force that does not depend on position nor on time. Later, after
Hubble discovered that the universe is expanding, Einstein
discarded this constant term, as the ``biggest blunder of his
life".

However, recent observations are indicating that a cosmological
constant may play an essential role in the cosmic expansion of the
Universe~\cite{Armendariz}. According to the standard cosmology
the total energy density of the Universe is dominated today by
both dark matter and dark energy densities. The dark energy
component in general is considered as a kind of vacuum energy
homogeneously distributed with negative pressure. The
observational data provide compelling evidence for the existence
of dark energy which dominates the present day Universe and
accelerates its expansion.

In principle, any matter content which violates the strong energy
condition and possesses a positive energy density and a negative
pressure, may cause the dark energy effect of repulsive
gravitation. So the main problem of modern cosmology is to
identify this form of dark energy that dominates the universe
today. Dark energy composed of just a cosmological term  $\Lambda$
is fully consistent with existing observational data. A
cosmological constant can be associated with  a time independent
dark energy density. Another candidate is the phantom matter,
whose energy density increases with the expansion of the
universe~\cite{Copeland}.

In general relativity one can also consider other gravitational
configurations where a cosmological constant, and even the phantom
matter, may play an important role. For example the presence of
the cosmological constant allows charged and non--charged black
hole geometries in 2+1 gravity~\cite{Teitelboim}. On the other
hand wormholes, as well as black holes, are an extraordinary
consequence of Einstein's equations of general relativity and,
during recent decades, there has been a considerable interest in
the field of wormhole physics. For instance, Lorentzian wormhole
configurations~\cite{Morris,Visser} may be supported by phantom
matter~\cite{Lobo,Zaslavskii,Cataldo} or by a cosmological
constant~\cite{Lemos,Roman,Delgaty}.

Two separate directions emerged: one relating to static Lorentzian
wormholes and the other concerned with time-dependent Lorentzian
ones~\cite{Teo,Kar,Lobo0}. In both cases the interest has been
focused on traversable wormholes, which have no horizons, allowing
two-way passage through them. However, most of the efforts are
directed to study static configurations which are traversable. The
most striking property of such a wormhole is the violation of
energy conditions. This implies that the matter supporting the
traversable wormholes is exotic~\cite{Morris,Visser}, which means
that it has very strong negative pressures, or even that the
energy density is negative, as seen by static observers.

One interesting aspect that we shall consider here is the
possibility of sustaining traversable wormhole spacetimes via
exotic matter made out of phantom energy. The latter is considered
to be a possible candidate for explaining the late time
accelerated expansion of the Universe~\cite{Cataldo15}. This
phantom energy has a very strong negative pressure and violates
the null energy condition, so becoming a most promising ingredient
to sustain traversable wormholes. Notice however that in this case
we shall use the notion of phantom energy in a more extended sense
since, strictly speaking, the phantom matter is a homogeneously
distributed fluid, and here it will be an inhomogeneous and
anisotropic fluid~\cite{Sushkov,Lobo,Cataldo}. It is interesting
to note that in Ref.~\cite{DeBenedictis} is considered the
possibility that the enormous negative pressure in the center of a
dark energy star may, in principle, imply a topology change,
consequently opening up a tunnel and converting the dark energy
star into a wormhole. It should be mentioned that the stability of
some specific wormhole static configurations also has been
considered~\cite{LemosL}.

The theoretical construction of wormholes is usually performed by
using the method where, in order to have a desired metric, one is
free to fix the form of the metric functions, such as the redshift
and shape functions, or even the scale factor for evolving
wormholes. In this way one may have a redshift function without
horizons, or with a desired asymptotic. Unfortunately, in this
case we can obtain expressions for the energy and pressure
densities which are physically unreasonable.

For constructing wormholes another method can be used. One may
impose conditions on the stress–-energy tensor threading the
wormhole configuration. For example, one can impose the so--called
self–dual condition $\rho=\rho_t=0$, on the matter supporting the
wormhole, where $\rho=T_{\alpha \beta}u^{\alpha}u^{\beta}$ and
$\rho_t=(T_{\alpha \beta}-\frac{1}{2}Tg_{\alpha
\beta})u^{\alpha}u^{\beta}$ are the energy density measured by a
static observer~\cite{Dadhich}. One can also prescribe the matter
content by specifying the equations of state of the radial and of
the tangential pressures or by choosing a more specific matter
field, such as, for example, a classical scalar field. Notice
that, as a simple model of phantom matter, one may consider a
classical scalar field with a negative kinetic energy term (the so
called ghost or phantom scalar field). Static~\cite{Ellis} and
non--static wormholes~\cite{SushkovA} were constructed with the
help of such a ghost scalar field. In all these here described
cases one can solve the Einstein field equations in order to find
all the metric functions.

In this paper we shall consider the radial and tangential
pressures of the matter threading the wormhole to obey barotropic
equations of state with constant state parameters coupled to a
cosmological constant. More specifically, we shall find all
evolving wormhole geometries which have their radial and
tangential pressures proportional to the energy density in the
presence of a cosmological constant.

The outline of the present paper is as follows: In Sec. II, we
briefly review some important aspects of evolving wormholes and
write the Einstein equations for matter configurations considered
here. In Sec. III, we obtain the general metrics which describe
evolving wormholes with anisotropic pressures obeying barotropic
equations of state with constant state parameters in the presence
of a cosmological constant. In Sec. IV, the properties of the
obtained wormhole geometries are studied and discussed.

\section{Evolving Lorentzian wormholes and the Einstein field equations}
The metric ansatz of Morris and Thorne~\cite{Morris} for the
spacetime which describes a static Lorentzian wormhole is given by
\begin{eqnarray}\label{4wormhole}
ds^2=-e^{2\Phi(r)}dt^2+\frac{dr^2}{1-\frac{b(r)}{r}}+r^2(d\theta^2+sin^2
\theta d \varphi^2),
\end{eqnarray}
where $\Phi(r)$ is the redshift function, and $b(r)$ is the shape
function since it controls the shape of the wormhole. In order to
have a wormhole the functions $b(r)$ and $\Phi(r)$ must satisfy
the general constraints discussed by Morris and Thorne in
Ref.~\cite{Morris}. These constraints provide a minimum set of
conditions which lead, through an analysis of the embedding of the
spacelike slice of~(\ref{4wormhole}) in a Euclidean space, to a
geometry featuring two asymptotically flat regions connected by a
bridge. Although asymptotically flat wormhole geometries have been
extensively considered in the literature, asymptotically anti--de
Sitter wormholes are also of particular interest~\cite{Barcelo}.

Now, the evolving wormhole spacetime may be obtained by a simple
generalization of the original Morris and Thorne
metric~(\ref{4wormhole}) to a time-dependent metric given by
\begin{eqnarray}\label{evolving wormhole}
ds^2=-e^{2\Phi(t,r)}dt^2+ \nonumber \\ a(t)^2 \left(
\frac{dr^2}{1-\frac{b(r)}{r}}+r^2(d\theta^2+sin^2 \theta d
\varphi^2)\right),
\end{eqnarray}
where $a(t)$ is the scale factor of the universe. Note that the
essential characteristics of a wormhole geometry are still encoded
in the spacelike section. It is clear that if $b(r) \rightarrow 0$
and $\Phi(t,r) \rightarrow 0$ the~(\ref{evolving wormhole}) metric
becomes the flat Friedmann-Robertson-Walker (FRW) metric and, as
$a(t) \rightarrow const$ and $\Phi(t,r)=\Phi(r)$ it becomes the
static wormhole metric~(\ref{4wormhole}).

Now we shall write the Einstein equations for the~(\ref{evolving
wormhole}) metric. In order to simplify the analysis and the
physical interpretation we now introduce the proper orthonormal
basis as
\begin{eqnarray}
ds^2= - \theta^{(t)}
\theta^{(t)}+\theta^{(r)}\theta^{(r)}+\theta^{(\theta)}\theta^{(\theta)}+
\theta^{(\varphi)}\theta^{(\varphi)},
\end{eqnarray}
where the basis one--forms $\theta^{(\alpha)}$ are given by
\begin{eqnarray}\label{tetrad}
\theta^{(t)}= e^{\Phi(t,r)} dt; \,\,\,\,\, \theta^{(r)}=
\frac{a(t) \, dr}{\sqrt{1-\frac{b(r)}{r}}}; \nonumber \\
\theta^{(\theta)}=a(t) \, r d \theta;  \,\,\,\,\,
\theta^{(\varphi)}= a(t) \, r \, sin \, \theta \,d \varphi.
\end{eqnarray}

In general, for the metric~(\ref{evolving wormhole}) one might
introduce a matter source described by an imperfect fluid, with
its four--velocity $u^{(\alpha)}$ oriented along the tetrad
$\theta^{(t)}$. Thus for the space--time~(\ref{evolving wormhole})
one might consider a stress-energy tensor of an imperfect fluid
containing the coefficients of bulk viscosity, shear viscosity and
heat conduction (for the considered metric we might have energy
flux in the radial direction), where all of these transport
coefficients might be functions of $t$ and/or of $r$. Of course in
this case the energy--momentum tensor has also non--diagonal
entries. However, as we have pointed out above, we shall use the
notion of phantom energy in a slightly more extended sense: We
shall consider this traditionally homogeneous and isotropic exotic
source to be generalized to an inhomogeneous and anisotropic
fluid, but still with a diagonal energy-momentum tensor (as is
usually considered in phantom cosmologies). This means that, for
the tetrad basis~(\ref{tetrad}), the only nonzero components of
the energy--momentum tensor $T_{(\mu)(\nu)}$ are the diagonal
terms $T_{(t)(t)}$, $T_{(r)(r)}$, $T_{(\theta)(\theta)}$ and
$T_{(\varphi)(\varphi)}$, which are given by
\begin{eqnarray}\label{EMT}
T_{(t)(t)}=\rho(t,r),
T_{(r)(r)}=p_r(t,r)=-\tau(t,r),\nonumber \\
T_{(\theta)(\theta)}=T_ {(\varphi)(\varphi)}=p_{_l}(t,r),
\end{eqnarray}
where the quantities $\rho(t,r)$, $p_r(t,r)$,
$\tau(t,r)(=-p_r(t,r))$, and
$p_{_l}(t,r)(=p_{\varphi}(t,r)=p_{\theta}(t,r))$ are respectively
the energy density, the radial pressure, the radial tension per
unit area, and the lateral pressure as measured by observers who
always remain at rest at constant $r$, $\theta$, $\varphi$. Thus
for the evolving spherically symmetric wormhole
metric~(\ref{evolving wormhole}) the Einstein equations with
cosmological constant $\Lambda$
\begin{eqnarray*}
R_{(\alpha)(\beta)}-\frac{R}{2} g_{(\alpha)(\beta)}=-\kappa
T_{(\alpha)(\beta)}-\Lambda g_{(\alpha)(\beta)}
\end{eqnarray*}
are given by
\begin{eqnarray}\label{00}
3 e^{-2\phi(t,r)} H^2+\frac{b^{\prime}}{a^2 r^2}=\kappa \rho(t,r)+\Lambda,  \\
\label{rr} - e^{-2\phi(t,r)}\, \left( 2\frac{\ddot{a}}{a}+ H^2
\right)- \frac{b}{a^2 r^3} + 2 e^{-2\phi(t,r)} H \frac{\partial
\phi}{\partial t}+ \\
\nonumber \frac{2}{r^2 a^2} (r-b ) \frac{\partial \phi}{\partial r}=\kappa p_r(t,r)-\Lambda, \\
\label{thetatheta} -e^{-\phi(t,r)} \left(2 \frac{\ddot{a}}{a}+ H^2
\right) + \frac{b-r b^{\prime}} {2 a^2 r^3}+ \\ \nonumber 2
e^{-\phi(t,r)} H \frac{\partial \phi}{\partial t} +
\frac{1}{2a^2r^2} (2r-b-rb^{\prime}) \frac{\partial \phi}{\partial
r} + \\ \nonumber \frac{1}{a^2r} (r-b)\left( \left(\frac{\partial
\phi}{\partial r} \right)^2+\frac{\partial^2 \phi}{\partial
r^2}\right)=\kappa
p_{_l}(t,r)-\Lambda,  \\
2 e^{-\phi(t,r)} \sqrt{\frac{r-b(r)}{r}} \,  \frac{\partial
\phi}{\partial r} \, \dot{a}=0, \label{thetaphi}
\end{eqnarray}
where $\kappa=8 \pi G$, $H=\dot{a}/a$, and an overdot and a prime
denote differentiation $d/dt$ and $d/dr$ respectively.

It is direct to see that, for the diagonal energy--momentum
tensor~(\ref{EMT}), Eq.~(\ref{thetaphi}) gives some constraints on
relevant metric functions which separate the wormhole solutions
into two branches: one static branch given by the condition
$\dot{a}=0$ and another non-static branch for $\partial
\phi/\partial r=0$. This latter condition implies that the
redshift function can only be a function of $t$, i.e.
$\phi(t,r)=f(t)$ so, without any loss of generality, by rescaling
the time coordinate we can set $\phi(t,r)=0$.

It is interesting to note that the static branch with $\phi(r)
\neq const$ allows one to have isotropic pressures for the matter
threading the wormhole (see for example~\cite{Lobo}). As we shall
see below, since for the non-static branch we must fulfill the
condition $\phi(t,r)=0$, the matter pressures cannot be isotropic
at all.

In what follows we shall restrict our discussion to evolving
wormhole geometries, so we must consider the non-static branch.
Thus, as we stated above, without any loss of generality we shall
require $\phi(t,r)=0$. By using the conservation equation
$T^\mu_{\,\,\,\,\nu;\mu}=0$, we have that
\begin{eqnarray}\label{CE1}
\frac{\partial \rho}{\partial t}+H (3 \rho+p_r+2p_{_l})=0, \\
\frac{2(p_{_l}-p_r)}{r}=\frac{2(p_{_l}+\tau)}{r}=\frac{\partial
p_r}{\partial r}. \label{CE2}
\end{eqnarray}
From these equations we see that for an isotropic pressure, i.e.
$p_{_l}=p_r=p$, we have to require $\partial p_r/\partial r=0$, so
the pressure will depend only on time $t$, obtaining the standard
cosmological conservation equation $\dot{\rho}+3 H (\rho+p)=0$. If
we want to study wormholes with pressures depending on both
variables $t$ and $r$, we must consider only anisotropic
pressures, thus requiring $ p_r \neq p_{_l}$.

In the following we are interested in studying matter sources
threading the wormhole described by barotropic equations of state
with constant state parameters. Thus we shall require that
\begin{eqnarray} \label{C00}
\tau(t,r)=-p_r(t,r)=-\omega_r \, \rho(t,r), \nonumber \\
p_{_l}(t,r)=\omega_{_l} \, \rho(t,r),
\end{eqnarray}
where $\omega_r$ and $\omega_{_l}$ are constant state parameters.
Clearly, the requirement~(\ref{C00}) with
$\omega_r=\omega_{_l}=\omega$ allows us to connect the evolving
wormhole spacetime~(\ref{evolving wormhole}) with the standard FRW
cosmologies, where the isotropic pressure density is expressed as
$p=\omega \rho$, with constant state parameter $\omega$.

Now, with the help of the conservation equations~(\ref{CE1})
and~(\ref{CE2}) we can easily solve the Einstein
equations~(\ref{00})--(\ref{thetaphi}). Note that in these
equations the cosmological constant is not present, so we can
solve the field equations following the procedure of
Ref.~\cite{Cataldo}. From the structure of these conservation
equations we see that one can write the energy density in the form
$\rho(t,r)=\rho_t(t)\rho_r(r)$. Thus from the conservation
equations~(\ref{CE1}) and~(\ref{CE2}) we obtain
\begin{eqnarray}\label{rhot}
\rho_t(t)=C_1 a^{-(3+\omega_r+2\omega_{_l})}, \\
\rho_r(r)=C_2 r^{2(\omega_{_l}-\omega_r)/\omega_r} \label{rhor},
\end{eqnarray}
where $C_1$ and $C_2$ are integration constants, obtaining for the
full energy density
\begin{equation}\label{rho}
\rho(t,r)=C \, r^{2(\omega_{_l}-\omega_r)/\omega_r} \,
a^{-(3+\omega_r+2\omega_{_l})},
\end{equation}
where we have introduced a new constant $C$ in order to redefine
$C_1$ and $C_2$.

Now, by subtracting Eqs.~(\ref{rr}) and~(\ref{thetatheta}), and
using the full energy density~(\ref{rho}), we obtain the
differential equation
\begin{equation}
\frac{\kappa (\omega_{_l}-\omega_r)\, C \,
r^{2(\omega_{_l}-\omega_r)/\omega_r}}{a^{(3+\omega_r+2\omega_{_l})}}=\frac{3b-r
b^{\prime}}{2a^2r^3}.
\end{equation}
It is straightforward to see that in order to have a solution for
the shape function $b=b(r)$ we must impose the constraint
\begin{equation}\label{constraint}
\omega_r+2\omega_{_l}+1=0
\end{equation}
on the state parameters $\omega_r$ and $\omega_{_l}$, thus
obtaining for the shape function
\begin{equation}\label{br}
b(r)=C_3 r^3-\kappa \, C \, \omega_r \, r^{-1/\omega_r},
\end{equation}
where $C_3$ is a new integration constant. Notice that the
constraint~(\ref{constraint}) implies that the radial and
tangential pressures are given by
\begin{equation}\label{pr pl}
p_r=\omega_r \rho, \,\,\, p_{_l}= -\frac{1}{2} \, (1+\omega_r)
\rho,
\end{equation}
so the energy density and pressures satisfy the following
relation:
\begin{equation}
\rho+p_r+2 p_{_l}=0.
\end{equation}

Now, from Eqs.~(\ref{00}),~(\ref{rho}),~(\ref{br}) and taking into
account the constraint~(\ref{constraint}) we obtain the following
master equation for the scale factor:
\begin{equation}\label{scf}
3 H^2=-\frac{3C_3}{a^2}+\Lambda.
\end{equation}
In the following paragraph, with the help of this equation, we
shall determine the scale factor, and then the specific form of
the energy density~(\ref{rho}).

Lastly, note that there is another branch of spherically symmetric
solutions to Eqs.~(\ref{00})--(\ref{thetatheta}). By adding these
equations and taking into account Eqs.~(\ref{C00}) and~(\ref{rho})
we obtain the equation
\begin{equation}\label{acceleration}
6\frac{\ddot{a}}{a}-2\Lambda=-\kappa (1+\omega_r+2\omega_{_l})\,C
\, r^{2(\omega_{_l}-\omega_r)/\omega_r} \,
a^{-(3+\omega_r+2\omega_{_l})}.
\end{equation}
Now, if we take $\omega_r=\omega_{_l}=\omega$, we obtain a
differential equation for $a(t)$ and $\Lambda$ which has a first
integral given by
\begin{equation}\label{acc}
3 \left( \frac{\dot{a}}{a} \right)^2+\frac{Const}{a^2}= \kappa C
a(t)^{-3(1+\omega)}+\Lambda.
\end{equation}
This differential equation is the standard Friedmann equation for
a FRW universe filled with an ideal fluid (with $p(t)=\omega
\rho(t)$) and a cosmological constant. The term $Const/a^2$ is
related to the curvature term while the energy density scales as
$\rho \sim a^{-3(1+\omega)}$.

It is worth noticing that, if we take $1+\omega_r+2\omega_{_l}=0$
(see Eq.~(\ref{constraint})), we obtain from
Eq.~(\ref{acceleration}) the differential equation
$3\ddot{a}/a=\Lambda$ which has a first integral of the form of
the master equation~(\ref{scf}).

\section{Wormhole solutions}
Now we shall study the specific wormhole configurations which
should be allowed by the master equation~(\ref{scf}). The case
$\Lambda=0$ (which implies that $a(t) = \sqrt{-C_3} \, t+ Const$)
was considered before in Ref.~\cite{Cataldo} so we shall restrict
ourselves here to the case $\Lambda \neq 0$. To start with, we
shall consider first the static case of a de Sitter wormhole.

\subsection{The $C_3=0$ case: The de Sitter wormhole}
First, let us consider the case $C_3=0$ in Eq.~(\ref{scf}). In
this case the solution is given by
\begin{equation}\label{SFE}
a(t)= a_0 e^{\pm \sqrt{\Lambda/3} \, t},
\end{equation}
and is allowed only for $\Lambda >0$. Thus the
metric~(\ref{evolving wormhole}) takes the form
\begin{eqnarray}\label{WHLambda}
ds^2 = dt^2+ a^2_0 e^{2\pm \sqrt{\Lambda/3} \, t}  \times \\
\nonumber\left( \frac{dr^2}{1+\kappa C \omega_r \,
r^{-(1+1/\omega_r)}}+r^2(d\theta^2+sin^2 \theta d
\varphi^2)\right),
\end{eqnarray}
describing contracting and expanding wormholes. If we choose the
plus sign in Eq.~(\ref{SFE}) we obtain an inflationary scale
factor, which gives an exponential expansion for an inflating
wormhole. This kind of wormholes was considered before by Roman in
Ref.~\cite{Roman}; however, in this work the resulting properties
are associated with a stress-energy tensor of a general form,
without imposing a specific equation of state among the energy
density and the pressures. In this case, for an inflating
wormhole, we obtain from Eq.~(\ref{C00}) that the anisotropic
pressures take the form~(\ref{pr pl}) where the energy density is
given by
\begin{eqnarray}\label{rhoLambda}
\rho(t,r) \sim \,\frac{r^{-(1+3\omega_r)/\omega_r}}{e^{2
\sqrt{\Lambda/3} \, t}}.
\end{eqnarray}
Since we want to study a wormhole configuration sustained via
exotic matter made out of phantom energy, we are interested in the
range $\omega_r < -1$ for the main state parameter of the
solution. Clearly, in this case for $r \rightarrow \infty$ we have
$\rho \rightarrow 0$, thus at spatial infinity this wormhole looks
like a de Sitter universe.

Now we shall rewrite this wormhole solution in a more appropriate
form. From the condition for the throat that the $r$--coordinate
has a minimum at $r_0$, i.e. $g^{-1}_{rr}(r_0)=0$, we obtain for
the integration constant
$C=-\frac{r_0^{(1+\omega_r)/\omega_r}}{\kappa \omega_r}$, yielding
for the shape function and the energy density
\begin{eqnarray}\label{whstatic}
b(r)=r_0\left(\frac{r}{r_0}\right)^{-1/\omega_r},
\kappa
\rho(t,r)=-\frac{(r/r_0)^{-(1+3\omega_r)/\omega_r}}{\omega_r r^2_0
e^{2 \sqrt{\Lambda/3} \, t}}, \nonumber \\
\end{eqnarray}
respectively. Thus the metric ~(\ref{WHLambda}) takes the form
\begin{eqnarray}\label{WHWHLambda}
ds^2 = dt^2+ a^2_0 e^{2\pm \sqrt{\Lambda/3} \, t}  \times \\
\nonumber\left( \frac{dr^2}{1-(r/r_0)^{-(1+\omega_r)/\omega_r}}+
\right. \left. r^2(d\theta^2+sin^2 \theta d \varphi^2)
\frac{}{}\right).
\end{eqnarray}

The radial coordinate $r$ has a range that increases from a
minimum value at $r_0$, corresponding to the wormhole throat, to
infinity. From Eqs.~(\ref{pr pl}), (\ref{whstatic})
and~(\ref{WHWHLambda}) we can see that for $\omega_r < -1$ the
wormhole solution is an asymptotically de Sitter universe with a
matter content having a radial pressure with a phantom equation of
state and everywhere positive energy density. Note that for
$\omega_r>0$ we also have a wormhole solution with an
asymptotically de Sitter universe, but in this case the energy
density is negative everywhere. For $\Lambda=0$ we have a static
wormhole solution considered before in Ref.~\cite{Lobo} (see also
Ref.~\cite{Cataldo}).

\subsection{The $C_3 \neq 0$ case}
Let us now explore the features of the general solution for the
above found evolving Lorentzian wormholes supported by the
considered anisotropic matter and cosmological constant. From the
master equation~(\ref{scf}) we can see that $H^2>0$ only for the
following combinations of the parameters $C_3$ and $\Lambda$:
$C_3>0$ and $\Lambda >0$, or $C_3<0$ and $\Lambda >0$, or $C_3<0$
and $\Lambda <0$. The general solution for Eq.~(\ref{scf}) with
$\Lambda \neq 0$ is given by
\begin{eqnarray}
a(t)=\frac{\sqrt{3}}{6\sqrt{\Lambda}} \left[ \frac{9 C_3+C^2_4
e^{\pm 2 \sqrt{\Lambda/3} \, t}}{C_4 e^{\pm \sqrt{\Lambda/3} \,
t}} \right],
\end{eqnarray}
where $C_4$ is a new integration constant. For the cases
enumerated before, the solution may be rewritten in the form shown
in table~\ref{tabla15}, where $\phi_0$ is a constant.
\begin{table}[h]
\begin{tabular}{|c|c|c|}
  \hline
  $a_{_{\Lambda}}(t)$ & $C_3 \neq 0$ & $\Lambda \neq 0$\\
  \hline
  $\sqrt{\frac{3 C_3}{\Lambda}} \cosh \left( \sqrt{\frac{\Lambda}{3}} \, t +\phi_0 \right)$ & $>0$ & $>0$  \\
\hline
 $\sqrt{\frac{-3 C_3}{\Lambda}} \sinh \left( \sqrt{\frac{\Lambda}{3}} \, t +\phi_0 \right)$ & $<0$ & $>0$  \\
\hline
  $\sqrt{\frac{3 C_3}{\Lambda}} \sin \left( \sqrt{\frac{-\Lambda}{3}} \, t +\phi_0 \right) $ &  $<0$ & $<0$   \\
  \hline
\end{tabular}
\caption{The table shows the possible scale factors derived from
Eq.~(\ref{scf}) and corresponding to the cases $C_3 \neq 0$ and
$\Lambda \neq 0$. \label{tabla15}}
\end{table}

Thus the solution for $C_3 \neq 0$ is given by the
metric~(\ref{evolving wormhole}) with $\Phi(t,r)=0$, the shape
function~(\ref{br}) and the scale factors indicated in
Table~\ref{tabla15}. However, it must be noted that in this
solution the radial coordinate may be rescaled in order to absorb
the integration constant $C_3$. In this case we obtain the final
metric given by
\begin{eqnarray}\label{evolving wormhole final}
ds^2=-dt^2+ a_{_{\Lambda}}^2(t) \times \nonumber\\ \left(
\frac{dr^2}{1-k r^2+\kappa C \omega_r r^{-(1
+\omega_r)/\omega_r}}+r^2(d\theta^2+sin^2 \theta d
\varphi^2)\right), \nonumber \\
\end{eqnarray}
where $k=1$ for $C_3>0$, $k=-1$ for $C_3<0$. Note also that we can
include in the rescaled metric~(\ref{evolving wormhole final}) the
case considered in the previous subsection making $k=0$ for
$C_3=0$. We summarize in Table~\ref{tabla915} all possible scale
factors for the found wormhole solutions~(\ref{evolving wormhole
final}).
\begin{table}[h]
\begin{tabular}{|c|c|c|}
  \hline
  $a_{_{\Lambda}}(t)$ & $k$ & $\Lambda$\\
  \hline
  $ const$ & $0$ & $0$  \\
  \hline
  $ t+ const$ & $-1$ & $0$  \\
 \hline
  $ a_0 e^{\pm \sqrt{\Lambda/3} \, t}$ & $0$ & $>0$  \\
\hline
  $\sqrt{\frac{3 }{\Lambda}} \cosh \left( \sqrt{\frac{\Lambda}{3}} \, t +\phi_0 \right)$ & $1$ & $>0$  \\
\hline
 $\sqrt{\frac{3 }{\Lambda}} \sinh \left( \sqrt{\frac{\Lambda}{3}} \, t +\phi_0 \right)$ & $-1$ & $>0$  \\
\hline
  $\sqrt{\frac{-3 }{\Lambda}} \sin \left( \sqrt{\frac{-\Lambda}{3}} \, t +\phi_0 \right) $ &  $-1$ & $<0$   \\
  \hline
\end{tabular}
\caption{The table shows all the possible scale factors for the
general solution~(\ref{evolving wormhole final}) of an evolving
Lorentzian wormhole with the radial tension and the tangential
pressure having barotropic equations of state~(\ref{C00}) with
constant state parameters. Here we have included the case $k=-1$,
$\Lambda = 0$ considered before in
Ref.~\cite{Cataldo}.\label{tabla915}}
\end{table}

In this case evolving wormholes have a matter source with the
anisotropic pressures given by Eq.~(\ref{pr pl}) where now the
energy density has the form
\begin{eqnarray}\label{rhoLambdaL}
\rho(t,r) =\frac{C \,
r^{-(1+3\omega_r)/\omega_r}}{a^2_{_{\Lambda}}(t)},
\end{eqnarray}
with $a_{_{\Lambda}}$ given in Table~\ref{tabla915}.

Now we shall rewrite the wormhole metric~(\ref{evolving wormhole
final}) in a more appropriate form. From the condition
$g^{-1}_{rr}(r=r_0)=0$, we obtain for the integration constant
\begin{eqnarray}
C=\frac{(kr^2_0-1)}{\kappa \omega_r} \,
r_0^{(1+\omega_r)/\omega_r},
\end{eqnarray}
yielding for the shape function and the metric component $g_{rr}$
\begin{eqnarray}\label{whevolving}
b(r)=r_0\left(\frac{r}{r_0}\right)^{-1/\omega_r}
\,\,\,\,\,\,\,\,\,\,\,\,\,\,\,\,\,\,\,\,\,\,\,\,\,\,\,\,\,\,
\nonumber
\\ +kr^3_0 \left(\frac{r}{r_0}\right)^3
\left(1-\left(\frac{r}{r_0}\right)^{-(1+3\omega_r)/\omega_r}\right),
\nonumber \\ a^2_{_\Lambda}(t) g_{rr}^{-1}= 1-
\left(\frac{r}{r_0}\right)^{-(1+\omega_r)/\omega_r}
\,\,\,\,\,\,\,\,\,\,\,\,\,\,\,\,\,\,\,\,\,\,\,\,\,\,\,\,\,\,
\nonumber
\\ -kr^2_0 \left(\frac{r}{r_0}\right)^2
\left(1-\left(\frac{r}{r_0}\right)^{-(1+3\omega_r)/\omega_r}\right),
\end{eqnarray}
respectively. This implies that the wormhole throat is located at
$r_0$. In this case the energy density of the matter threading the
wormhole takes the following form:
\begin{eqnarray}\label{rrr}
\kappa \rho(t,r)=\frac{kr^2_0-1}{r^2_0 \omega_r a_{_\Lambda}(t)^2}
\left(\frac{r}{r_0}\right)^{-(1+3\omega_r)/\omega_r}.
\end{eqnarray}
Clearly, in order to have an evolving wormhole we must require
$\omega_r < -1$ or $\omega_r >0$ (in both of these cases, in the
$g_{rr}$ metric component, $(1+\omega_r)/\omega_r>0$ and
$(1+3\omega_r)/\omega_r>0$), implying that the phantom energy can
support the existence of evolving wormholes in the presence of a
cosmological constant. As before, for a matter made of phantom
energy with $\omega_r<-1$, we have that at spatial infinity its
energy density vanishes. Thus the wormhole configurations with
$k=-1,1$ (or equivalently $C_3 \neq 0$) are asymptotically vacuum
$\Lambda$-Friedmann models with open and closed topologies
respectively with the corresponding scale factors of the
table~\ref{tabla915}. For $\omega_r>0$ we have the same behavior.

\section{Properties of the solutions and discussion}
Now we shall consider some characteristic properties of the above
found evolving wormhole geometries. Essentially, as we shall see,
these wormholes coupled to a cosmological constant have the same
properties as have the evolving wormholes with $\Lambda=0$
discussed in Ref.~\cite{Cataldo}, since the cosmological constant
directly controls the behavior of the scale factor $a(t)$ and not
the shape function $b(r)$ of the spacelike section of the
metric~(\ref{evolving wormhole}). However it must be noticed that
in this case we have a richer variety of topologies since now we
can consider evolving wormhole configurations with $k=0$ and
$k=1$, while only the case $k=-1$ for $\Lambda=0$. Effectively,
for example, from the metric~(\ref{evolving wormhole final}) we
can see that for wormholes supported by phantom matter at spatial
infinity ($r\rightarrow \infty$) we have the following asymptotic
metric:
\begin{eqnarray}\label{wormhole asintotico}
ds^2 \approx -dt^2+ \nonumber\\a_{_{\Lambda}}^2(t)  \left(
\frac{dr^2}{1-k r^2}+r^2(d\theta^2+sin^2 \theta d
\varphi^2)\right).
\end{eqnarray}
This metric has slices $t=$ const which are 3--spaces of constant
curvature: open for $k=-1$, flat for $k=0$ and closed for $k=1$.
This implies that the asymptotic metric~(\ref{wormhole
asintotico}) is foliated with spaces of constant curvature.

Now the presence of the cosmological constant of course indicates
that these wormholes are not asymptotically flat. If we calculate
the Riemann tensor for the metric~(\ref{evolving wormhole final}),
we find that its independent non--vanishing components are
\begin{eqnarray}\label{CRT}
R_{(\theta)(r)(r)(\theta)}=R_{(\varphi)(r)(r)(\varphi)}=
\nonumber \\
\frac{\dot{a}^2+k+\frac{1}{2}\kappa C (1+\omega_r) \, r^{-(1+3
\omega_r)/\omega_r}}{a^2}, \nonumber \\
R_{(\varphi)(\theta)(\theta)(\varphi)}=\frac{\dot{a}^2+k-\frac{1}{2}\kappa
C \omega_r \, r^{-(1+3 \omega_r)/\omega_r}}{a^2}, \nonumber \\
R_{(t)(\theta)(t)(\theta)}=R_{(t)(\varphi)(t)(\varphi)}=R_{(t)(r)(t)(r)}=
\frac{\ddot{a}}{a}. \,\,\,\,\,\,\,\,\,\,
\end{eqnarray}
From these expressions and the scale factors given in
Table~\ref{tabla915} we see that at spatial infinity, i.e. $r
\rightarrow \infty$, these Riemann tensor components do not vanish
for a wormhole with $\omega_r <-1$ or $\omega_r >0$, except for
the case $\Lambda=0$ as stated in Ref.~\cite{Cataldo}. In
conclusion, regardless of the energy density~(\ref{rhoLambdaL})
vanishing for $r \rightarrow \infty$, the Riemann tensor
components do not vanish, due to the presence of the cosmological
constant $\Lambda$.

Let us now consider the energy conditions. It is well known that,
in all cases, the violation of the weak energy condition (WEC) is
a necessary condition for a static wormhole to exist. In general,
for the energy--momentum tensor~(\ref{EMT}), WEC reduces to the
following inequalities:
\begin{eqnarray}
\rho(t,r)\geq0, \,\,\, \rho(t,r)+p_r(t,r)\geq0, \nonumber \\
\rho(t,r)+p_{_l}(t,r)\geq0,
\end{eqnarray}
which may be rewritten by using the expressions~(\ref{C00})
and~(\ref{constraint}) as follows:
\begin{eqnarray}\label{WEC}
\rho(t,r)\geq0, \,\,\, (1+\omega_r) \, \rho(t,r) \geq0, \nonumber \\
(1-\omega_r) \, \rho(t,r)  \geq0.
\end{eqnarray}
Thus, for the found gravitational configurations (for which we
must require $\omega_r<-1$ or $\omega_r>0$ in order to have a
wormhole), Eqs.~(\ref{rrr}) and~(\ref{WEC}) imply that for:
\begin{itemize}
\item $k=0$ or $k=-1$ and $\rho>0$, we must require $\omega_r<-1$,
so the WEC is always violated. Note that this is independent of
the value of the cosmological constant.
\item $k=1$ (in this case
$\Lambda >0$) and $\rho>0$, we may require $\omega_r<-1$ for
$r^2_0 <1$ (and the WEC is always violated) or require
$0<\omega_r<1$ for $r^2_0
>1$ (and the violation of WEC is avoided). Unfortunately this
latter case must be ruled out for the consideration of evolving
wormhole configurations as we shall see below.
\end{itemize}

Note that for the case $k=1$ the value $r_0=1$ implies that
$\rho=0$ and $b(r)=r^3$, so we have in this case a homogeneous and
isotropic closed $\Lambda$--FRW cosmology.

In order to dilucidate whether or not the shape of the wormhole is
maintained, we must check the fulfillment of the flaring out
condition, which is given by
\begin{eqnarray}\label{FOC}
\frac{d^2\bar{r}}{d\bar{z}^2}=\frac{\bar{b}-\bar{b}^\prime r}{2
\bar{b}^2}=\frac{b-b^\prime r}{2 a(t_0)b^2}>0,
\end{eqnarray}
where $\bar{r}=a_\Lambda(t_0) r$, $\bar{b}(\bar{r})=a_\Lambda(t_0)
b(r)$ (for a detailed discussion see Ref.~\cite{Cataldo}). Taking
into account the shape function $b(r)$ of Eq.~(\ref{whevolving})
we obtain
\begin{eqnarray}
\frac{d^2\bar{r}}{d\bar{z}^2}=
-\left(\frac{r}{r_0}\right)^{1/\omega_r} \times
\,\,\,\,\,\,\,\,\,\,\,\,\,\,\,\,\,\,\,\,\,\,\,\,\,\,\,\,\,\,
\,\,\,\,\,\,\,\,\,\, \nonumber \\
\frac{2D\omega_r
r^3(r/r_0)^{1/\omega_r}+(1+\omega_r)r_0(kr_0^2-1)}{2\omega_r
(kr^3(r/r_0)^{1/\omega_r}+r_0(1-kr_0^2))^2}.
\end{eqnarray}
Clearly if $k=0,-1$ this constraint is satisfied for the entire
range of the radial coordinate $r$. If $k=1$ this constraint is
always violated for $r_0 >1$, and for $r_0<1$ it may be satisfied.
Thus we can have wormholes with $k=1$ only for $r_0<1$ and then
WEC is violated.

On the other hand, from Eq.~(\ref{acceleration}) and the
constraint~(\ref{constraint}) we conclude that the expansion of
the wormhole is accelerated due to the presence of the
cosmological constant. So this family of evolving wormholes,
supported by an anisotropic phantom energy, may expand with
acceleration for $k=1,0,-1$. More specifically, from the form of
the scale factors of Table~\ref{tabla915}, we conclude that for
the cases with $\Lambda>0$ the wormhole expands forever, and for
$\Lambda <0$ the wormhole expands to a maximum value and then
recolapses.


The shape of a wormhole is determined by $b(r)$ as viewed, for
example, in an embedding diagram in a flat $3$--dimensional
Euclidean space $R^3$. To construct in our case such a diagram of
a wormhole we can follow the procedure described in
Ref.~\cite{Cataldo}. In order to have a good embedding we must
impose the condition $b(r)\geq 0$. Clearly from
Eq.~(\ref{whevolving}) we see that for $k=0$ and $k=1$ the shape
function $b(r)$ is positive for all $r>r_0$. For $k=-1$ the shape
function is positive for $r_0 <r< r_{max}$ where $r_{max}=r_0
\left(1+\frac{1}{r_0^2}\right)^{\omega_r/(1+3\omega_r)}$.

Let us now consider the tidal forces experienced by a traveller
while crossing this kind of wormholes. We recall that the tidal
acceleration must not exceed one Earth gravity, i.e. $g_{\oplus}=
9.8$ m$/s^2$, in order for wormhole travel to be at all convenient
for human beings~\cite{Morris}. The calculus of these tidal forces
may be simplified by introducing a new orthonormal reference frame
which moves at a constant speed $v$ with respect to observers who
always remain at rest at constant $r$, $\theta$ and $\varphi$ (see
Ref.~\cite{Cataldo} and also cites therein).

Thus, for the generic metric~(\ref{evolving wormhole}) (with
$\Phi(t,r)=0$), and by considering the size of the traveller's
body (i.e. head to feet) to be $\sim 2$ (m), the Riemann tensor
components in this new basis are constrained to be
\begin{eqnarray}\label{R1}
|R_{{\hat{1}^\prime} \hat{0}^\prime \hat{1}^\prime \hat{0}
}|=\left|\frac{\ddot{a}}{a} \right|\leq \frac{g_{\oplus}}{c^2
\times 2 \, m} \simeq \frac{1}{(10^{8} m)^2},
\end{eqnarray}
and
\begin{eqnarray}\label{R2}
|R_{{\hat{2}^\prime} \hat{0}^\prime \hat{2}^\prime
\hat{0}^\prime}|= |R_{{\hat{3}^\prime} \hat{0}^\prime
\hat{3}^\prime \hat{0}^\prime}|=  \nonumber \\ \left |\gamma^2
\frac{\ddot{a}}{a}-\frac{\gamma^2 \beta^2}{2 a^2 r^3} \left( 2
\dot{a}^2 r^3-b+r b^{\prime}\right) \right|\leq \nonumber \\
\frac{g_{\oplus}}{c^2 \times 2 \, m} \simeq \frac{1}{(10^{8}
m)^2}, \nonumber \\
\end{eqnarray}
where $\gamma=1/\sqrt{1-\beta^2}$ with $\beta=v/c$.

In this case the radial tidal constraint~(\ref{R1}) can be
regarded as directly constraining the acceleration of the
expansion of the wormhole, while the lateral tidal
constraint~(\ref{R2}) can be regarded as constraining the speed
$v$ of the traveller while crossing the wormhole.

In particular, the evolving wormholes considered in this paper
evolve with scale factors shown in Table~\ref{tabla915}. This
implies that the expansion is not accelerated (i.e. $\ddot{a}=0$)
only for the cases $\Lambda= 0$, thus satisfying the
constraint~(\ref{R1}). Now, by taking into account
Eq.~(\ref{acceleration}), we conclude that for all $\Lambda \neq
0$ cases of Table~\ref{tabla915}, the radial tidal
constraint~(\ref{R1}) implies the following constraint on the
cosmological constant:
\begin{eqnarray}
\left |\frac{\Lambda}{3} \right| \leq \frac{g_{\oplus}}{c^2 \times
2 \, m} \simeq \frac{1}{(10^{8} m)^2}.
\end{eqnarray}
Note that this constraint on the cosmological constant is valid
for all values of $k=-1,0,1$.

Now, by using the Einstein equation~(\ref{thetatheta}) with
Eqs.~(\ref{constraint}) and~(\ref{acceleration}), the lateral
tidal constraint~(\ref{R2}) may be rewritten as follows:
\begin{eqnarray*}
\left|\frac{\Lambda}{3}\gamma^4 -\frac{1}{2}
 \gamma^2 \beta^2 (1+\omega_r) \kappa \rho \right| \leq
\frac{g_{\oplus}}{c^2 \times 2 \, m} \simeq \frac{1}{(10^{8}
m)^2}.
\end{eqnarray*}
Since constraint~(\ref{R1}) implies that the cosmological constant
$\Lambda \leq 10^{-16}$ m$^{-2}$ and the motion of the traveller
be non--relativistic (i.e. $v<<c$, $\gamma \approx 1$), the above
constraint may be rewritten as
\begin{eqnarray}\label{R215}
\left|\frac{1}{2} \left(\frac{v}{c}\right)^2 (1+\omega_r) \kappa
\rho \right| \leq \frac{g_{\oplus}}{c^2 \times 2 \, m} \simeq
\frac{1}{(10^{8}
m)^2}. \nonumber \\
\end{eqnarray}
Thus the lateral tidal constraint~(\ref{R2}) can be regarded more
exactly as constraining both the speed $v$ of the traveller and
the energy density of the matter threading the wormhole. By taking
into account the expression for the energy density~(\ref{rrr}), we
may rewrite Eq.~(\ref{R215}) as follows:
\begin{eqnarray}\label{constraint3}
\left| \frac{v^2 (1+\omega_r)(kr^2_0-1)}{r^2_0 \omega_r
a_{_\Lambda}(t)^2}
\left(\frac{r}{r_0}\right)^{-(1+3\omega_r)/\omega_r}\right|
\lesssim g_{\oplus}.
\end{eqnarray}
As we can see, this constraint is more severe at the wormhole
throat, thus evaluating it at $r=r_0$ we obtain
\begin{eqnarray}\label{constraint315}
\left| \frac{v^2 (1+\omega_r)(kr^2_0-1)}{r^2_0 \omega_r
a_{_\Lambda}(t)^2} \right| \lesssim g_{\oplus}.
\end{eqnarray}
Thus we conclude that for wormholes with $k=-1,0,1$ and $\Lambda
\neq 0$ it is possible to fulfill the
constraint~(\ref{constraint315}) for some $t \geq t_{min}>0$. For
example for the case $k=0$ and $a(t)=a_0 e^{\sqrt{\Lambda/3} \,
t}$ we have that the constraint~(\ref{constraint315}) is saturated
for
\begin{eqnarray*}
t_{e_0}=\frac{1}{2 \sqrt{\Lambda/3}} \ln
\left(-\frac{v^2(1+\omega_r)}{g_{\oplus} r_0^2 \omega_r a_0^2}
\right),
\end{eqnarray*}
with $-\frac{v^2(1+\omega_r)}{g_{\oplus} r_0^2 \omega_r a_0^2}
>1$, thus for any $t \geq t_{e_0}>0$ the mentioned
lateral tidal constraint will be satisfied.

\section{Acknowledgements}
This work was supported by CONICYT through Grants FONDECYT N$^0$
1080530 and 1070306 (MC, SdC and PS), and by Direcci\'on de
Investigaci\'on de la Universidad del B\'\i o--B\'\i o (MC). SdC
also was supported by PUCV grant N$^0$ 123.787/2008 and P.S. and
P.M. by Universidad de Concepci\'on through DIUC Grant N$^0$
208.011.048-1.0.


\begin{thebibliography}{2}
\bibitem{Armendariz} C.~Armendariz-Picon, V.~F.~Mukhanov and P.~J.~Steinhardt, Phys.\
Rev.\ Lett.\  {\bf 85}, 4438 (2000); A.~I.~Arbab, Class.\ Quant.\
Grav.\  {\bf 20}, 93 (2003); M.~Giovannini, Int.\ J.\ Mod.\ Phys.\
A {\bf 22}, 2697 (2007).
\bibitem{Copeland} E.~J.~Copeland, M.~Sami and S.~Tsujikawa, Int.\ J.\ Mod.\ Phys.\ D
{\bf 15}, 1753 (2006)
\bibitem{Teitelboim} M.~Banados, C.~Teitelboim and J.~Zanelli,
Phys.\ Rev.\ Lett.\  {\bf 69}, 1849 (1992);  C.~Martinez,
C.~Teitelboim and J.~Zanelli, Phys.\ Rev.\  D {\bf 61}, 104013
(2000); M.~Cataldo, N.~Cruz, S.~del Campo and A.~Garcia, Phys.\
Lett.\  B {\bf 484}, 154 (2000); M.~Cataldo, Phys.\ Lett.\  B {\bf
529}, 143 (2002).
\bibitem{Morris} M.S. Morris and K.S. Thorne, Am.
J. Phys. {\bf 56}, 395 (1988); M.S. Morris, K.S. Thorne and U.
Yurtsever, Phys. Rev. Lett. {\bf 61}, 1446 (1988).
\bibitem{Visser} M. Visser, Lorentzian Wormholes: From Einstein to Hawking, (AIP,
New York, 1995); M. Visser, S. Kar, N. Dadhich Phys. Rev. Lett.
{\bf 90} 201102 (2003); N. Dadhich, S. Kar, S. Mukherjee and M.
Visser, Phys. Rev. D {\bf 65}, 064004 (2002).
\bibitem{Sushkov} S.~V.~Sushkov, Phys.\ Rev.\  D {\bf 71}, 043520
(2005).
\bibitem{Lobo} F.~S.~N.~Lobo, Phys.\ Rev.\  D {\bf 71}, 084011
(2005).
\bibitem{Zaslavskii} R.~Garattini and F.~S.~N.~Lobo, Class.\ Quant.\ Grav.\
{\bf 24}, 2401 (2007) ; F.~S.~N.~Lobo, Phys.\ Rev.\  D {\bf 71},
124022 (2005); O.~B.~Zaslavskii, Phys.\ Rev.\ D {\bf 72}, 061303
(2005); F.~Rahaman, M.~Kalam, M.~Sarker and K.~Gayen, Phys.\
Lett.\  B {\bf 633}, 161 (2006); P.~K.~F.~Kuhfittig, Class.\
Quant.\ Grav.\ {\bf 23}, 5853 (2006); P.~F.~Gonzalez-Diaz, Phys.\
Lett.\  B {\bf 632}, 159 (2006).
\bibitem{Cataldo} M. Cataldo, P. Labra\~na, S. del
Campo, J. Crisostomo and P. Salgado, Phys.\ Rev.\  D {\bf 78},
104006 (2008).
\bibitem{Lemos} J.~P.~S.~Lemos, F.~S.~N.~Lobo and S.~Quinet de Oliveira, Phys.\ Rev.\  D {\bf 68}, 064004
(2003).
\bibitem{Roman} T.A. Roman, Phys. Rev. D {\bf 47}, 1370 (1993).
\bibitem{Delgaty} M.~S.~R.~Delgaty and R.~B.~Mann, Int.\ J.\ Mod.\ Phys.\  D {\bf 4}, 231
(1995); B.~N.~Esfahani, Gen.\ Rel.\ Grav.\  {\bf 37}, 271 (2005);
S.~W.~Kim, Phys.\ Lett.\  A {\bf 166}, 13 (1992).
\bibitem{Teo} E. Teo, Phys. Rev. D {\bf 58},
024014 (1998); V.M. Khatsymovsky, Phys. Lett. B {\bf 429}, 254
(1998); P. K. F. Kuhfittig, Phys. Rev. D 67, 064015 (2003);
Tonatiuh Matos, D. Nunez Class. Quant. Grav. {\bf 23}, 4485
(2006); Mubasher Jamil, Muneer Ahmad Rashid, Electromagnetic field
around a slowly rotating wormhole arXiv: 0805.0966 [astro-ph].
\bibitem{Kar} S. Kar, Phys. Rev. D {\bf 49}, 862 (1994); S. Kar and D. Sahdev, Phys. Rev. D {\bf 53}, 722 (1996).
\bibitem{Lobo0} F.S.N. Lobo, Exotic solutions in General Relativity: Traversable wormholes
and 'warp drive' spacetimes, e-Print: arXiv:0710.4474 [gr-qc]; A.
V. B. Arellano and F. S. N. Lobo; Class. Quant. Grav. {\bf23},
5811 (2006); A. V. B. Arellano and F. S. N. Lobo; Class. Quant.
Grav. {\bf 23}, 7229 (2006).
\bibitem{Cataldo15} S.~Nojiri, S.~D.~Odintsov and S.~Tsujikawa,
Phys.\ Rev.\  D {\bf 71}, 063004 (2005); P.~F.~Gonzalez-Diaz,
Phys.\ Rev.\  D {\bf 68}, 021303 (2003); P.~F.~Gonzalez-Diaz,
Phys.\ Lett.\  B {\bf 586}, 1 (2004); M.~Cataldo, N.~Cruz and
S.~Lepe, Phys.\ Lett.\  B {\bf 619}, 5 (2005);  G.~Izquierdo and
D.~Pavon, Phys.\ Lett.\  B {\bf 633}, 420 (2006).
\bibitem{DeBenedictis} A.~DeBenedictis, R.~Garattini and F.~S.~N.~Lobo, arXiv:0808.0839 [gr-qc].
\bibitem{LemosL} J.~P.~S.~Lemos and F.~S.~N.~Lobo, Phys.\ Rev.\  D {\bf 78}, 044030
(2008); J.~A.~Gonzalez, F.~S.~Guzman and O.~Sarbach,
arXiv:0806.0608 [gr-qc]; J.~A.~Gonzalez, F.~S.~Guzman and
O.~Sarbach, arXiv:0806.1370 [gr-qc]; D.~H.~Correa, J.~Oliva and
R.~Troncoso, JHEP {\bf 0808}, 081 (2008).
\bibitem{Dadhich} N. Dadhich, S. Kar, S. Mukherjee and M. Visser, Phys. Rev. D {\bf 65},
064004 (2002); M.~Cataldo, P.~Salgado and P.~Minning, Phys.\ Rev.\
D {\bf 66}, 124008 (2002).
\bibitem{Ellis}  L.~A.~Gergely, Phys.\ Rev.\  D {\bf 65}, 127503 (2002); S.~A.~Hayward,
Phys.\ Rev.\  D {\bf 65}, 124016 (2002); H. Ellis, J. Math. Phys.
{\bf 14}, 104 (1973); K. A. Bronnikov, Acta Phys. Polonica B {\bf
4}, 251 (1973).
\bibitem{SushkovA} S.~V.~Sushkov and Y.~Z.~Zhang, Phys.\ Rev.\  D {\bf 77}, 024042
(2008); S.~V.~Sushkov and S.~W.~Kim, Gen.\ Rel.\ Grav.\  {\bf 36},
1671 (2004).
\bibitem{Barcelo} C.~Barcelo, L.~J.~Garay, P.~F.~Gonzalez-Diaz and G.~A.~Mena
Marugan, Phys.\ Rev.\  D {\bf 53}, 3162 (1996).
\end{thebibliography}
\end{document}